\documentclass[prb,twocolumn,superscriptaddress, floatfix]{revtex4-2}

\usepackage{graphicx,color}
\usepackage{amssymb}
\usepackage{amsmath}
\usepackage{amsfonts,amsthm} 
\usepackage{natbib}
\usepackage[colorlinks=true,linkcolor=blue,citecolor=blue,urlcolor=blue]{hyperref}
\usepackage{ulem}
\usepackage{physics}

\begin{document}

\title{Monopoles, Dirac strings and Magnetic Noise in Model Spin Ice}

\author{A. Huster Zapke}
\affiliation{ENS de Lyon, CNRS, Laboratoire de Physique, F-69342 Lyon, France}
\affiliation{Institute for Theoretical Physics, University of G\"ottingen, Friedrich-Hund-Platz 1, 37077 G\"ottingen, Germany}
\author{P. C. W. Holdsworth}
\affiliation{ENS de Lyon, CNRS, Laboratoire de Physique, F-69342 Lyon, France}
\affiliation{French American Center for Theoretical Science, CNRS, KITP, Santa Barbara, CA 93106-4030, USA}

\begin{abstract}
{Using the Dirac string formalism for monopoles we expose an extensive analogy between magnetic monopole excitations in the dumbbell model of spin ice and those of the vacuum. In both cases the Dirac strings are defined in the space-time of monopole trajectories which are simulated in spin ice using transition graphs between initial and final configurations. 
The Stanford experiment for Dirac monopole detection is reconstructed in spin ice using the fragmentation procedure. The setup is then extended to simulate magnetic noise experiments using stochastic dynamics. It is shown that the noise from the monopoles and their constraining strings can be separated and that the correlated signal over long times comes largely from the constraining strings rather than from the monopoles themselves. }
\end{abstract}

\maketitle


\section{Introduction}

Much has been written \cite{udagawaSpinIce2021} about frustrated magnets on a pyrochlore lattice  in terms of emergent gauge fields with $U(1)$ symmetry \cite{Isakov2004} and the consequent magnetic monopole quasi-particles in spin ice \cite{Harris1997,Bramwell2001,Bramwell2020,Castelnovo2008,Ryzhkin2005,Castelnovo2012}. These are taken to be analogues of Dirac monopoles with accompanying Dirac strings \cite{Dirac1931,Dirac1948}. The classical analogues of the Dirac strings can be identified in a special case in which a monopole path shows up against an ordered background \cite{Morris2009} but in general they remain ill-defined for arbitrary spin configurations \cite{Castelnovo2008,Jaubert2009,Jaubert2011,Castelnovo2012}. In this paper, after reviewing the Stanford experiment for monopole detection \cite{Cabrera1982} in a language suitable for adaptation we develop a precise definition of Dirac strings in spin ice as well as for the magnetic fields radiating from the monopoles. We show that, in close analogy with Dirac's theory, the strings are defined in the space-time of monopole trajectories rather than in individual microstates of the Gibbs ensemble of states. The space-time trajectories appear naturally for particles traversing a detector, allowing us to re-construct the Stanford experiment for the dumbbell model of spin ice \cite{Castelnovo2008,Brooks2014}. We further show that in both cases the Dirac string has a topological signature that strongly influences the measurement.
Using the same detector we are able to simulate magnetic noise experiments for a monopole fluid in analogy with recent
experiments and simulations on the spin ice material Dy$_2$Ti$_2$O$_7$  \cite{Dusad2019,Samarakoon2022,Morineau2025,billington2024}.  We confirm that the noise is due to the monopole activity but that the Dirac strings make a vital contribution that dominates the correlated signal at long times.

\begin{figure}[tp]
\includegraphics[trim=2.0cm 8.5cm 2.0cm 7cm,width = .7\linewidth]{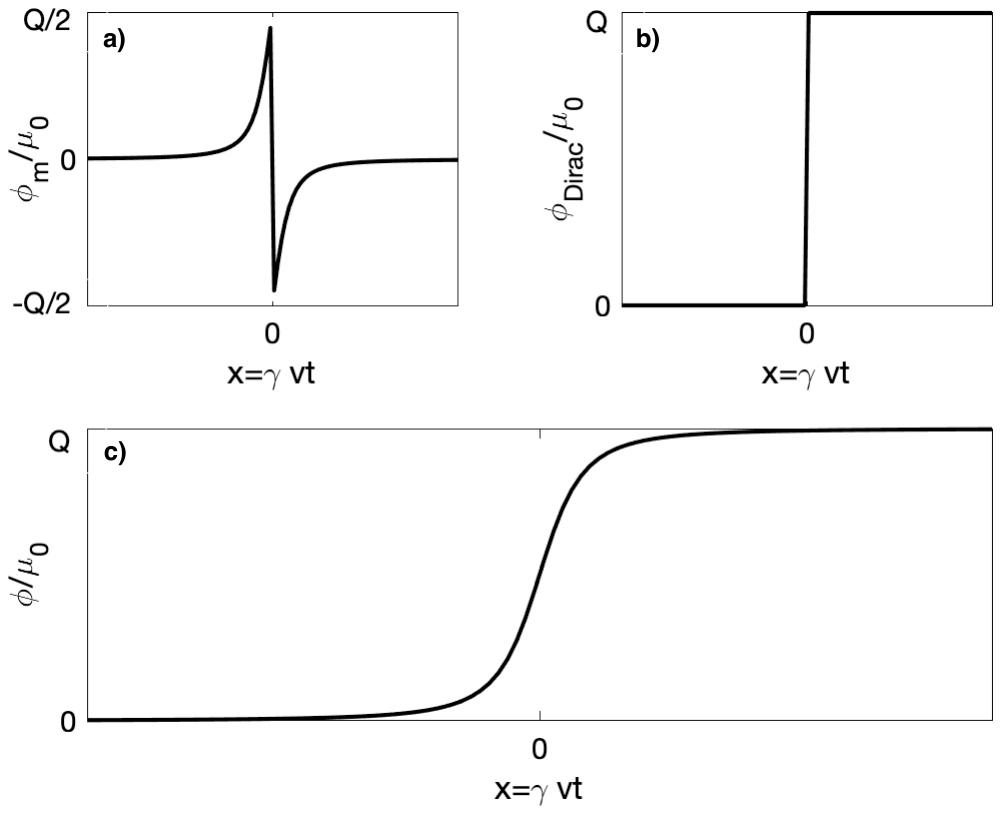}
 \caption{\label{Fig_Stanford} Flux change for the monopole through the Faraday coil. The total c) is the sum of the flux of the monopole a) plus that of the Dirac string b). }
\end{figure}

\section{The Stanford Experiment}

In Dirac's theory of deconfined magnetic charge \cite{Dirac1931} a monopole at position $\vec r(t)$ moves with velocity $\vec v$ and traces out a trajectory represented by the locus of points $\left[\vec s(t)\right]$ \cite{Dirac1948}. In his 1948 paper Dirac shows that the breakdown of Maxwell's equations is limited to a line singularity along $\left[\vec s(t)\right]$; the famous Dirac string. In Appendix (\ref{Maxwell}) we show that, by inclusion of the line singularity one can write down an expression for the magnetic field produced by a monopole that leaves Maxwell's equations 
unchanged and satisfied in all space:
 \begin{eqnarray}
 \vec B(\vec r\;^{\prime},t)&=&\mu_0 Q\left[\int^{\vec r(t)}_{-\infty}\delta(\vec s(t^{\prime})-\vec r\;^{\prime})d \vec s(t^{\prime})+ \frac{1}{4\pi (\vec x(t))^2}\hat{x}(t)\right]\nonumber \\ \label{Bfield}
 &=& \vec B_{Dirac} + \vec B_m,
 \end{eqnarray}
 where $Q$ is the monopole charge (positive charge for a north pole and negative for a south pole), $\mu_0$ is the permeability of free space, $d\vec s$ lies along the trajectory $\left[\vec s(t^{\prime})\right]$ for $-\infty<t^{\prime}<t$
 and $\vec x(t) =\vec r\;^{\prime}-\vec r(t)$ is the relative position of the observer ($\vec r\;^{\prime}$) and the particle ($\vec r(t)$). The first term is the line singularity associated with the Dirac string 
 and the second the field radiating from the monopole at $\vec r(t)$. 
 The modified field is divergence free as the Dirac string provides the flux radiating from the point charge.
 The monopole current is now replaced by the Dirac string dynamics. It is important to stress that this space-time formulation goes beyond magneto-statics as the string cannot be placed without prior knowledge of the particle trajectory. 
 
 
An integral part of Dirac's theory is that the Dirac string should be unobservable, unphysical even \cite{Dirac1948}. In particular, in a quantum interference experiment, the wave function of an electron of charge $-e$ whose trajectories encircle the singular line should incur a phase shift through the Aharanov-Bohm effect \cite{Aharanov1959}. The shift is predicted to be an integer times $2\pi$, leaving no trace of the string and leading to the quantification condition for the product of electron and monopole charge: $\mu_0eQ=nh$ where $h$ is Planck's constant. 

The formulation of eqn. (1) and eqns; (\ref{Max2}) which we use below does however provide a physical interpretation for the strings which is extremely useful for analogies with spin ice. In this case, although the string is unobservable the monopole is not and the termination of the string on the monopole has a determining effect on the the measurement. As proposed in the Stanford experiment \cite{Cabrera1982}, a monopole could be detected through the change in magnetic field flux $\phi=\int \vec B . d\vec S$ through a conducting coil enclosing a surface with elements $d\vec S$ normal to it. An electromotive force is generated via Faraday's law: $emf=-\frac{d\phi}{dt}$ (see Appendix (\ref{Maxwell})) driving a current $I$ which saturates in a superconducting coil due to the self-inductance $L$: $emf=-L\frac{dI}{dt}$ giving a one to one correspondence between $I$ and $\phi$.

Following the construction of ref. [\onlinecite{Cabrera1982}] we consider the ring to be circular, of radius $b$, lying in the plane perpendicular to $\hat{x}$ with centre at the origin and the particle to be travelling with uniform speed $v$ along $\hat{x}$.  At time $t$ its distance of approach as seen by the coil is $x=\gamma vt$ with $\gamma$  the relativistic Lorentz function. 
The monopole radiates flux isotropically so that its contribution, $\phi_m$ depends on the solid angle of approach and changes sign as the monopole passes through the coil:
\begin{eqnarray}
\phi_m(t)&=&\frac{\mu_0 Q}{2}\left(1-2\theta(t)+\frac{\gamma vt}{\left[(\gamma vt)^2+b^2\right]^{1/2}}\right) \nonumber \\
\phi_{Dirac}(t)&=&\mu_0Q\theta{(t)}\nonumber \\
\phi&=&\phi_{Dirac}+\phi_m, \label{flux}
\end{eqnarray}
where $\theta(x)$ is the Heaviside step function. In addition, the Dirac string terminates at the monopole so its contribution, $\phi_{Dirac}$ changes discontinuously (eqn.(\ref{flux})) as the particle passes through. The signal, $\phi(t)=\phi_{Dirac}+\phi_m$ corresponding to the passage of a monopole tethered by a string is illustrated in Fig. 1.
Although the string is undetectable at a point far from the monopole its end points are an integral part of the monopole signal and its presence introduces a topological constraint to the vacuum. As a consequence a signal from a passage through the coil is radically different from a near miss. For example, a trajectory that stalls in front of the the coil at $x=-\epsilon$ then leapfrogs over the coil to $x=\epsilon$ before continuing as before would record the signal in Fig. (1a) rather than the predicted signal, Fig. (1c).

Famously, a single event was recorded on 14th February 1982 that has all the characteristics of a Dirac monopole tethered by a Dirac string in the way described above. Sadly a second, confirming event has not so far been observed leaving the validity of Dirac's theory an open problem.

\begin{figure}[tp]
 \includegraphics[width = .9\linewidth]{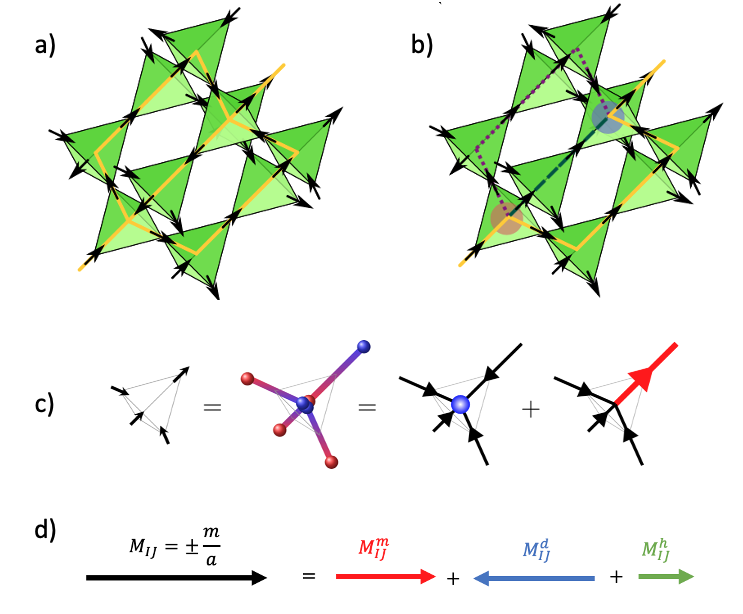}
 \caption{\label{Fig2--New} {\bf a)} A low energy spin ice configuration satisfying the ice-rules. Yellow lines show flippable loops. {\bf b)} A spin ice configuration with broken ice rules and magnetic monopoles. The blue and maroon spheres are north and south poles. The dashed, dotted and yellow lines between the monopole are possible return paths. {\bf c)} In the dumbbell model, spins are replaced by needles of magnetic flux terminating in magnetic charge at the tetrahedron centre. The needles of this isolated three in-one out tetrahedron (north pole) fragment into elements $M_{IJ}^m$ and $M^d_{IJ}$ (see text). {\bf d)} Flux element $M_{IJ}$ fragmented into elements in the monopole, gauge and harmonic sectors.}
\end{figure}

\section{Magnetic Monopoles in Spin Ice}

In spin ice materials the magnetic ions sit on the sites of the pyrochlore lattice defining corner sharing tetrahedra (see Fig. 2). It is a face centred cubic structure of tetrahedral unit cells of four sites. Alternatively a cubic cell of side $a_c$, containing four unit cells and 16 spins can be used. Spins are separated by distance $a_s=\frac{a_c}{2\sqrt{2}}$ and tetrahedron centres, which form a diamond lattice are separated by $a=\frac{\sqrt{3}a_c}{4}$. The spins are constrained to point along the body centres of the cubic structure towards, or away from the tetrahedron centres. The lowest energy configurations satisfy the ice rules \cite{Pauling1935,Harris1997} with two spins pointing in and two out of every tetrahedron. These form an extensive low energy band of (Pauling) states with loops of spins that can be flipped taking the system to different states within the band, as illustrated in Fig. 2a. Monopole quasi-particles \cite{Castelnovo2008,Ryzhkin2005} are localised in the centres of tetrahedra in which the ice rules are broken, with three spins in - one out or three out - one in (Fig 2b). In the nearest neighbour model of spin ice \cite{Bramwell1998}, the spins of unit length map to elements of a lattice field. In the band of Pauling states, which are  degenerate ground states the field is the curl of a gauge potential with $U(1)$ symmetry \cite{Isakov2004} and the monopoles are topological defects in this field. However, the moments $m$ of spin ice materials Ho$_2$Ti$_2$O$_7$ and Dy$_2$Ti$_2$O$_7$ are large, approximately 10 Bohr magnetons per spin so that the dipole interaction must also be taken into account \cite{denHertog2000}. Thanks to the remarkable symmetry properties of the pyrochlore lattice \cite{denHertog2000,Isakov2005} the ground state degeneracy of the nearest neighbour model remains virtually \cite{Melko2001} intact on inclusion of the full dipole interaction but the topological defects, the magnetic monopoles acquire real Coulomb interactions between them \cite{Castelnovo2008,Kaiser2019}. 

The magnetic monopole approximation to spin ice is captured in full by the dumbbell model \cite{Castelnovo2008} in which the spins $\vec S_i$ on sites $i$ of the pyrochlore lattice are replaced by needle-like bar magnets extending between tetrahedron centres $I$ and $J$. The magnetic flux is channelled along the needles which terminate at north and south poles - dumbbells of magnetic charge $q=\pm m/a$. A two-in two out state has total charge zero on each tetrahedron and a monopole carries charge $Q=2q=\pm 2m/a$ (Fig. 2c). In this model the Pauling states are degenerate by construction and monopoles interact via a Coulomb interaction with energy above the ground state given by %
\begin{equation}
{\cal{H}}={u(a)\over{2}}\sum_{I\ne J} \left({a\over{r_{IJ}}}\right)\hat{n}_I\hat{n}_I - \mu \sum_I \hat{n}_I^2,
\label{eq2}
\end{equation} 
%
where $u(a)={\mu_0 Q^2\over{4\pi a }}$ is the nearest neighbour Coulomb energy scale for a pair of monopoles, $\hat{n}_I=0,\pm 1,\pm 2$ is a monopole number operator and $r_{IJ}$ the distance between the centre of tetrahedron  $I$ and $J$.
The spin ice problem is therefore equivalent to lattice Coulomb fluid in the grand ensemble \cite{Castelnovo2008,Jaubert2009,Raban2019,Kaiser2019}, allowing creation and annihilation events with chemical potential for monopole and double monopole creation $\mu$ and $\mu_2=4\mu$ respectively. 
For dysprosium titanate $\mu=-4.35$ K, $u(a)\approx 3$ K and $a=4.33$ \AA \cite{Kaiser2019}.

A field theoretic description of the problem \cite{Brooks2014} can be constructed by treating the needles as elements of a lattice field $M_{IJ}=-M_{JI}$ that satisfy properties of both the emergent field of the nearest neighbour model and the real magnetic flux of the system of coupled dipoles \cite{Brooks2014}. The latter dictates that a spin pointing into a tetrahedron carries a negative magnetic flux. 

The lattice field can be broken up or fragmented into its constituent parts; divergence full, gauge and harmonic components, via a lattice Helmholtz decomposition \cite{Brooks2014,Lhotel2020}
\begin{equation}
M_{IJ}=M_{IJ}^m+M_{IJ}^d+M_{IJ}^h, \label{3comp}
\end{equation}
where the $M_{IJ}^m$ constitute the monopoles. They have an associated Gauss' law for each tetrahedron $I$
\begin{equation}
\sum_{J=1,4} M_{IJ}^m=-Q_I,\label{divf}
\end{equation}
where $Q_I$ is the magnetic charge on site $I$ and the sum is over the four neighbours $J$. The minus sign, optional for the emergent field, is imposed here so that magnetic flux flows into a north pole of positive charge, ensuring that the flux of magnetic field radiates outward from it. Of the remaining terms $M_{IJ}^d$ provides closed loops of flux and are the lattice curl of a vector potential while the harmonic terms $M_{IJ}^h$ \cite{Bramwell2017,Museur2024} provide a constant background and capture a finite ferromagnetic moment. The emergence of continuous symmetry from discrete objects is illustrated by the decomposition: while $M_{IJ}=\pm m/a$ the three elements can take on continuous values constrained only by the total available flux (see Fig 2d). For the rest of the paper, for simplicity we will set $M^h_{IJ}=0$ and work in zero external field. 

The fragmentation procedure has been highly successful for the interpretation of partially ordered phases \cite{Lefrancois2019,Cathelin2020,Pearce2022,Museur2024} where one fragment is ordered and another is disordered. However, the principle can be applied to any spin ice configuration including those of monopole fluids. The simplest situation, illustrated in Fig. 2c is that of a tetrahedron hosting an isolated monopole, far from all others. The four elements hosting the north pole fragment in units of $\frac{Q}{2}$:
\begin{eqnarray}
  [-1,-1,-1,1]&=&[-\frac{1}{2},-\frac{1}{2},-\frac{1}{2},-\frac{1}{2}]\\ \nonumber
  &+&[-\frac{1}{2},-\frac{1}{2},-\frac{1}{2},\frac{3}{2}],\label{frag-simple}
\end{eqnarray}
where the first set are the $[M^m]$ satisfying eqn. (\ref{divf}) and the second set, satisfying Kirchoff's current law are the $[M^{d}]$. 

Beyond this tetrahedron, or in situations where monopoles interact, the field distributions are complex but can be calculated using the efficient real space iterative algorithm proposed by Slobinsky {\it et. al.} \cite{Slobinsky2019}. We have used their algorithm to isolate the $M_{IJ}^m$ for generic configurations (see Appendix \ref{Frag-App}). In Fig. (3) we show the resulting coarse grained fields  generated for a neutral pair of monopoles. The simulation is for a system of $L^3$ cubic cells ($L=8$) with periodic boundaries. The monopoles are separated by $L/2$ in units of $a_c$. 
The figure shows that the relevant field from an isolated monopole does indeed radiate magnetic flux  isotropically to or from its singular source.

\begin{figure}[tp]
 \includegraphics[width = .9\linewidth]{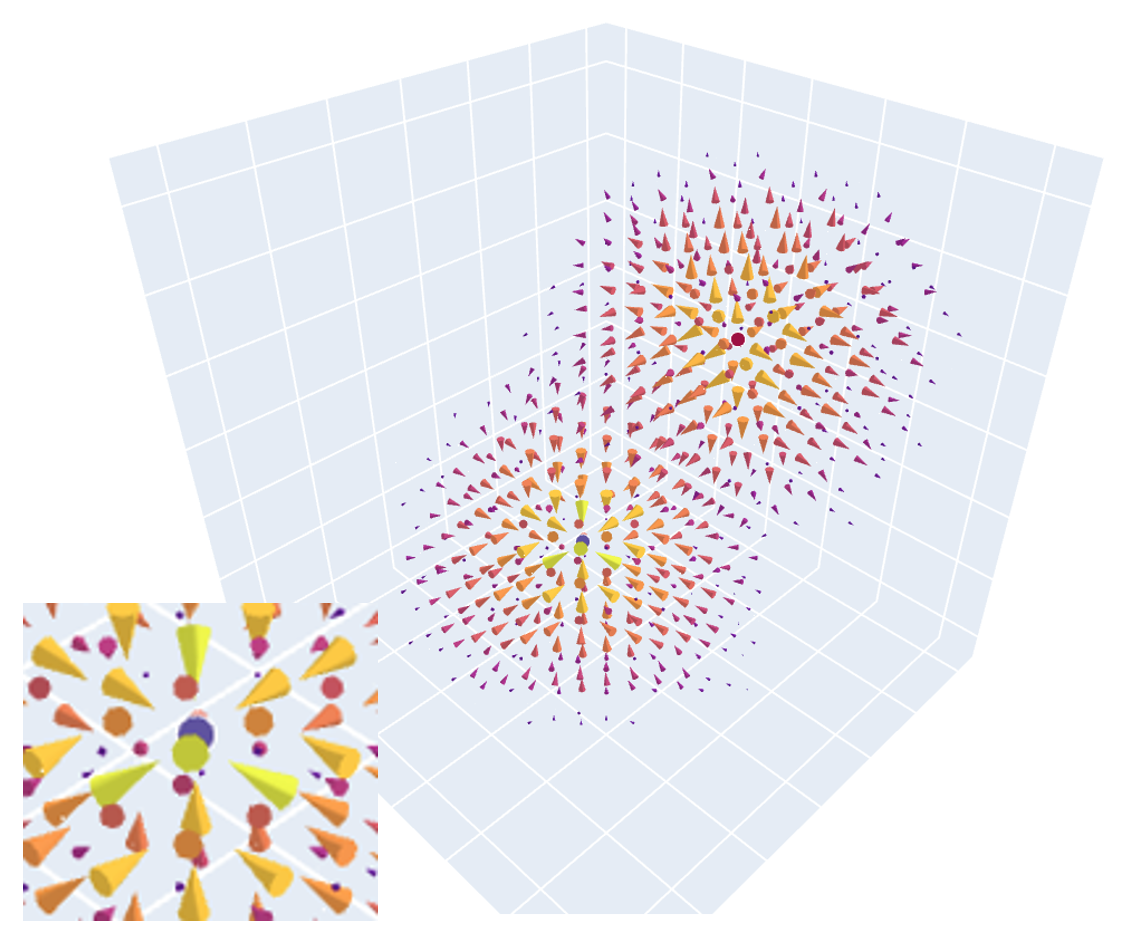}
 \caption{\label{flux-phim} {\bf{(Main)}} The monopole part of the coarse grained magnetic flux $M_{IJ}^m$ for a neutral pair of monopoles in a simulation of size $L=8$ separated by $L/2$. The north pole is blue, the south pole maroon. Each arrow is the average of, on average the four field values and field strength is indicated by arrow size on a logarithmic scale. Magnetic flux $M_{IJ}^m$ flows into the north pole. {\bf{(Lower left)}} Zoom in onto the north pole. }
\end{figure}

\section{Dirac strings and magnetostatics}

The network of available paths for monopole evolution, illustrated in Fig. 2b is often referred to as a Dirac string network \cite{Castelnovo2008,Jaubert2009,Jaubert2011,Castelnovo2012}. This definition requires some modification. As seen above, the Dirac string is intimately related to the monopole trajectory so that a single microstate cannot yield information concerning its position. To illustrate this consider the spin configuration in Fig. 2b. 
Without knowledge of the initial configuration there is no way of knowing if the north pole arrived at its present position by flipping spins along the dashed path, the dotted path or the yellow path. Adding the initial configuration it immediately becomes apparent that it followed the dashed path and that the Dirac string must lie along it. The network of needles carrying magnetic flux of $Q/2$ therefore offer a network of possible paths along which a Dirac string could lie, but no more can be said from a single microstate. 


It is clear from the above that, just as in the vacuum spin ice Dirac strings can only be identified through the particle's history. Although there is no time in equilibrium statistical mechanics, temporal evolution can be simulated by associating it with a given sequence of chosen microstates. The space-time of electrodynamics can then be simulated in spin ice by a transition graph \cite{Rokhsar1988} which subtracts the initial configuration from the final configuration following this sequence. Taking the initial configuration, $[M]_i$  to be a Pauling state and the final configuration, $[M]_f$ to be one in which a monopole pair has been created and the north pole stepped through a succession of sites, the transition graph yields a uniquely defined Dirac string
\begin{eqnarray}
[M]_{Dirac}=-\left([M]_f-[M]_i\right).\label{Dirac-s}
\end{eqnarray}
The field elements of the graph are of strength $Q$ along the trajectory \cite{Castelnovo2012} and zero elsewhere and the chosen sign orients them in a direction that evacuates the magnetic flux flowing into the north pole. An imprint of this protocol can be made in the special case in which a single monopole trajectory is observed against an ordered background, in spin ice through neutron scattering \cite{Morris2009} and through electron holography \cite{Dhar2021}  and in artificial spin ice \cite{Mengotti2011}. Simulations of the electron holography also show evidence of Dirac strings in disordered systems using transition graphs as outlined here.

We can now calculate the magnetic fields associated with the transition graph. In this model, magnetic fields are also channeled along the needles so that magnetic field and magnetic intensity elements can be defined:
%
\begin{equation}
B_{IJ}=\mu_0(M_{IJ}+H_{IJ}) \label{Bfield}.
\end{equation}
The magnetic intensity elements $H_{IJ}$ are related to the $M^m_{IJ}$ through the divergence free condition on $\vec B$ which imposes that  $\div. \vec M=-\div \vec H$ \cite{Castelnovo2008}. Putting this on the lattice and noting that spin ice materials are insulators, it follows that $M_{IJ}^m=-H_{IJ}$. Setting the harmonic term to zero for simplicity one then finds  that 
\begin{equation}
    B_{IJ}=\mu_0M^d_{IJ} \label{curlfree},
\end{equation}
ensuring  that the on-lattice magnetic field satisfies the divergence free condition.

\begin{figure}[tp]
 \includegraphics[width = .32\linewidth]{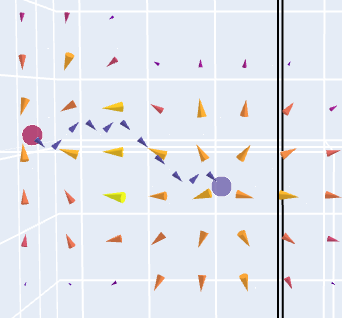}
 \includegraphics[width = .34\linewidth]{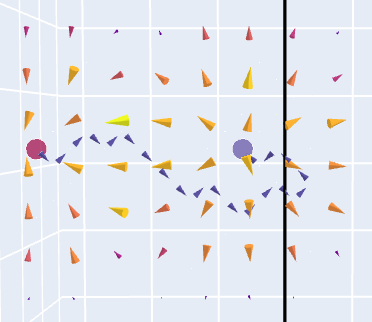}
 \includegraphics[width = .6\linewidth]{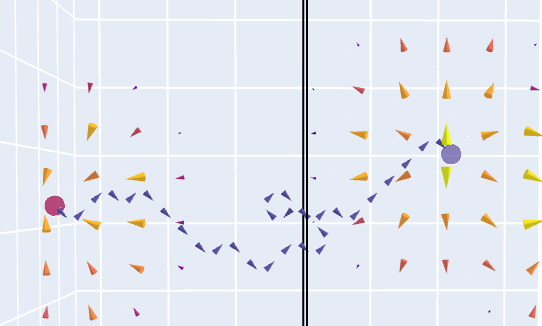}
 \caption{\label{Fig_Faraday} {Magnetic fields, eqn.(\ref{B-monopole}) for three stages of the monopole trajectory from left to right along $\hat{x}$, $L=8$. Images show the $x-y$ plane. Blue circle north pole, maroon south pole. Orange and yellow arrows show two dimensional projections of the magnetic fields from the monopoles from coarse graining over columnar cells of cross section $\Delta^2$ in the $x-y$ plane and $L$ in the $\hat{z}$ direction, $\Delta=0.8a_c$ giving 100 cells. Intensities are on a logarithmic scale. Blue arrows, individual elements of  the magnetic field along the Dirac string of strength $\mu_0Q-\mu_0|M_{IJ}^m|$. Black lines show position of the Faraday coil lying in the plane perpendicular to the $\hat{x}$ axis. }} 
\end{figure}

\begin{figure}[tp]
 \includegraphics[width = .6\linewidth]{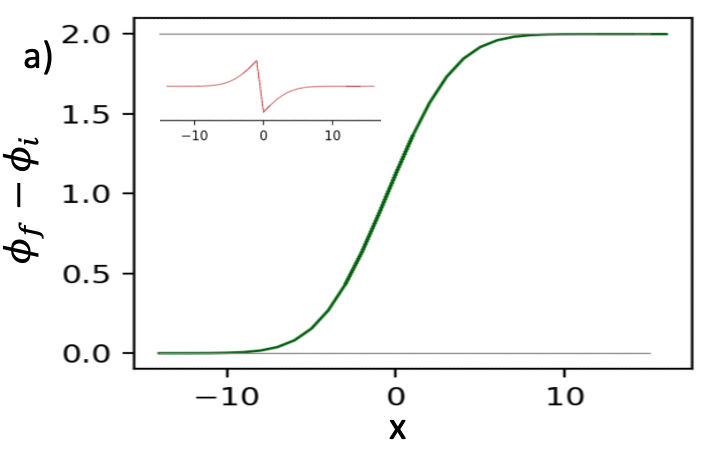}
 \includegraphics[width = .62\linewidth]{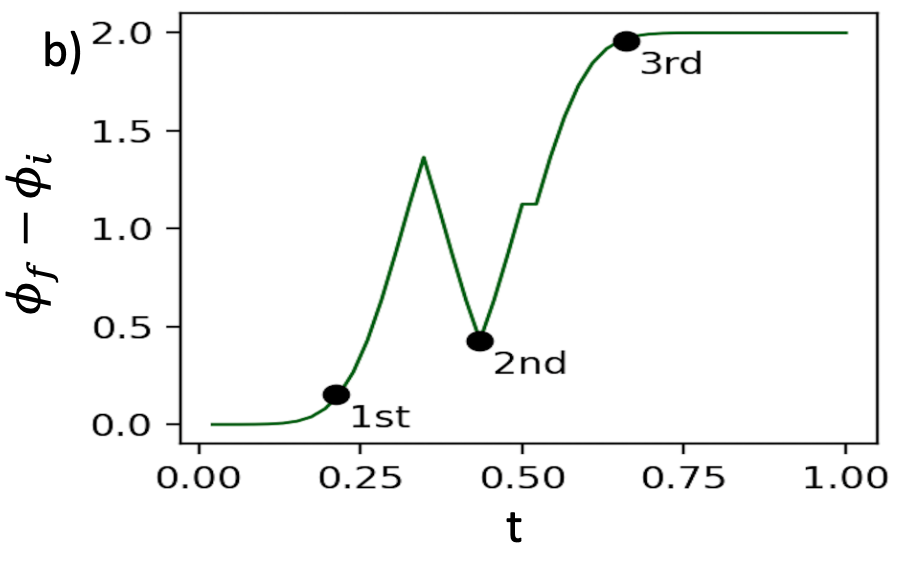}
 \caption{\label{Stanford-flux} {\bf a)} Flux difference through the coil (eqn. (\ref{phidiff})) in units of $\mu_0\left(\frac{Q}{2}\right)$ for the north pole {\it vs.} distance $x$ in units of $a_c/4$, the step distance along the $\hat{x}$ axis. Inset, evolution of the flux $\phi_m$ from the monopole ($[B]_m$ eqn.(\ref{B-monopole})). {\bf b)} Flux difference {\it vs} stepping time in units of total time. $1^{st}$, $2^{nd}$, $3^{rd}$ are for the three times of Fig. \ref{Fig_Faraday}. } 
\end{figure}

\begin{figure}[tp]
 \includegraphics[width = .62\linewidth]{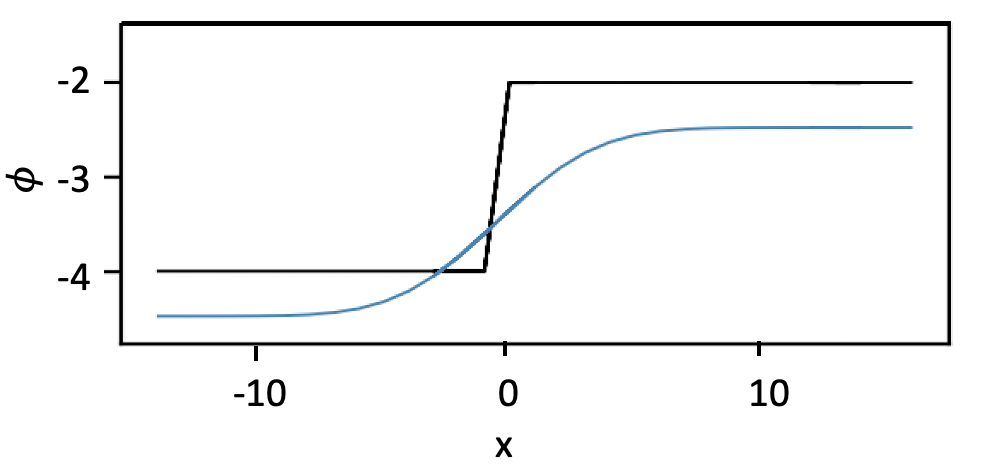}
 \caption{\label{Stanford-abs} Flux through the coil in units of $\mu_0\left(\frac{Q}{2}\right)$ for the north pole {\it vs.} distance $x$ in units of $a_c/4$, the step distance along the $\hat{x}$ axis. Blue: magnetic field flux $\phi$ (first term of eqn. (\ref{phidiff})). Black: local field flux $\phi_T$ (eqn.(\ref{phiT})). The blue data is shifted vertically for clarity.} 
\end{figure}

Fragmenting the final configuration of the sequence, $[M]_f=[M^m]_f+[M^d]_f$ and using eqns. (\ref{Dirac-s}) and (\ref{curlfree}) we can calculate the difference in magnetic field configurations associated with the transition graph:
\begin{eqnarray}
[B]_f - [B]_i &=&-\mu_0\left([M]_{Dirac}+[M^m]_f\right) \nonumber \\
&=&[B]_{Dirac}+[B]_m,\label{B-monopole}
\end{eqnarray}
where the first and second terms are the magnetic fields from the Dirac string and the monopole, in complete analogy with eqn. (1). 

A north pole (three spins in - one out) steps between sites by flipping one of the in spins to out. If $I$ is the initial and $J$ the final position of the monopole, the element $M_{IJ}$ flips from $-(\frac{Q}{2})$ to $\frac{Q}{2}$ producing an element of $[M]_{Dirac}$ of amplitude $Q$ flowing away from it (eqn.(\ref{Dirac-s})) and an element of $[B]_{Dirac}$ of amplitude $\mu_0Q$ flowing towards it. Field distributions given by eqn. (\ref{B-monopole}) at different moments of the trajectory are illustrated in Fig. (\ref{Fig_Faraday}), showing both the singular field from the monopole and the Dirac string. Magnetic field radiates outward from the north pole and the Dirac string term, $[B]_{Dirac}$ delivers the flux necessary to assure that the lattice field $[B]$ is everywhere divergence free.

\section{The Stanford Experiment in Spin Ice}

We can use this formalism to construct a monopole detection experiment for the dumbbell model. 
As illustrated in Fig. (\ref{Fig_Faraday}), a numerical Faraday coil is placed in the centre of the simulation cell, lying in the $\hat{y}-\hat{z}$ plane perpendicular to the $\hat{x}$ axis ($[100]$ cubic axis of the pyrochlore crystal). The plane of the  coil passes through the pyrochlore sites (spin positions) at $\vec x=0$ and cuts the needles spanning between tetrahedron centres $I,J$ on either side of the plane. 


As the magnetic field is confined to the needles  the lattice field elements, $B_{IJ}$ are scalars with units of flux. The flux through the coil is then just the sum of the field strengths passing through the needles cut by the coil. The flux change as the final state evolves is
\begin{equation} 
\phi_f - \phi_i=\left[\sum_{\langle I,J\rangle} B_{IJ}\right]_f - \left[\sum_{\langle I,J\rangle} B_{IJ}\right]_i, \label{phidiff}
\end{equation}
where $I$ is on the left of the coil and $J$ the right. Defining $I$ and $J$ in this way ensures that the flux calculated is that passing from left to right. 

The  flux difference with distance travelled by the north pole is shown in Fig. \ref{Stanford-flux}a. It changes smoothly with distance by a factor of $\mu_0 Q$ as the monopole steps through the coil. This is a clear monopole detection signal in complete analogy with with Fig. 1, confirming that the transition graph approach correctly simulates the space-time in which the monopole and Dirac string are predicted to exist \cite{Dirac1948}. Also shown is the flux change from the monopole alone, $\phi_m$, which is the contribution to eqn. (\ref{phidiff}) from $[B]_m$ in eqn. (\ref{B-monopole}). This flux increases as the coil is approached before jumping singularly to negative values and increasing back to zero for large $x$, again in close analogy with the monopole contribution to Fig. 1. 

One difference between the transition graph and the time evolution of the Dirac monopole is that here, even when stepped from left to right the trajectory is not ballistic as the pole encounters  constraints and stochastic choices along its path. As a consequence stepping time and distance are not equivalent.  This is illustrated in Fig. \ref{Stanford-flux}b were the flux change is given as a function of stepping time. The irregular jumps correspond to the enforced back tracking that can be observed in Fig. \ref{Fig_Faraday}. The trajectory also contains a step where the particle does not move, as could occur in a simulation with stochastic Monte Carlo dynamics.

Finally, in Fig. (\ref{Stanford-abs}) we show the flux through the coil without subtraction of the initial flux. Although the scale now depends on the total spin configuration the same change of flux can be observed. This shows that although the spin configuration consists of a forest of flux elements these contain an imprint of the monopole and its Dirac string which are exposed in the transition graph. Also shown is the local field flux
\begin{equation}
    \phi_T=\mu_0\left[\sum_{\langle I,J\rangle} M_{IJ}\right]_f,\label{phiT}
\end{equation}
from the elements of eqn.(\ref{3comp}). It shows a discrete spin flip from ``in'' to ``out'' when the monopole jumps through the plane of the Faraday coil. This is analogous  to the Dirac string contribution to the monopole signal (Fig.(1b)).

\section{A magnetic monopole noise experiment}

Recent magnetic noise experiments \cite{billington2024,Morineau2025,Samarakoon2022,Dusad2019} have highlighted the highly correlated spin dynamics of spin ice materials in the low and intermediate temperature ranges. The noise signal has been  interpreted in terms of monopole dynamics with correlations coming from Dirac strings \cite{Dusad2019}.  However, field distributions in materials and realistic models of spin ice \cite{Sala2012,Kirschner2018} are far more complex than those of the dumbbell model and detailed analysis has shown that some features of the strong correlations are non-universal effects specif to spin ice, rather than to the generic problem of magnetic monopoles and Dirac strings \cite{Morineau2025,Samarakoon2022,billington2024}. It therefore seems important to generate noise that is unequivocally from the fluctuating magnetic fields of magnetic monopoles and Dirac strings and compare with experimental results. 


%


To this end we have extended our protocol to measure such a signal during a finite temperature simulation using a single spin flip Monte Carlo algorithm. Equation (\ref{Dirac-s}) can be extended to define Dirac strings for a finite concentration of monopoles in a fluid configuration. A system, starting in a given Pauling state or vacuum $[M]_i$ is put in contact with a thermal reservoir at temperature $T$. Monopoles are created and move in the stochastic series of selected microstates, creating a set of Dirac strings. As the particles are deconfined and the strings tensionless \cite{Castelnovo2012}, the final configuration $[M]_f$ become uncorrelated with the initial state as thermal equilibrium is reached. Beyond this point the transition graph becomes a complex soup of strings terminating in monopoles and closed flux loops of strength $\mu_0Q$ \cite{Benton2012}. As each tetrahedron can accommodate $0$, $1$, or $2$ such flux tubes, the strings can cross and intersect with each other leading to budding and exchange of partners, characteristics that are surely shared with Dirac strings in the vacuum. As a consequence, in this regime the transition graph of the actual state with the initial ground state is of less interest than for the Stanford experiment. However, the topological properties of the vacuum ensure that a monopole passing through the coil drags a well defined Dirac string with it, for example along the dashed path in Fig. (\ref{Fig2--New}b). As well as the direct passage of monopoles, the signal will also be influenced by monopole creation, annihilation \cite{Klyuev2019} and movement near the coil due to the long range nature of the magnetic fields they produce.

We have used a sample with $L=4$ with periodic boundaries and with parameters for Dy$_2$Ti$_2$O$_7$. The simulations were run at 1 Kelvin, beginning from thermal equilibrium and the Monte Carlo time per spin was scaled to $t_{Nyq}=0.1$ mili-second as in [\onlinecite{Dusad2019}], which is the smallest (Nyquist) time interval. Run times were 50000 Monte Carlo steps per spin. The configurations were fragmented \cite{Slobinsky2019} and the flux through the Faraday coil recorded at each step, keeping track of the magnetic field flux $\phi$ ($\phi_f$ of eqn. (\ref{phidiff})), the flux from the monopoles only, $\phi_m$ (the contribution to eqn.(\ref{phidiff}) from  $[B]_m$ in eqn. (\ref{B-monopole})) and the total flux, eqn.(\ref{phiT}) from the local fields. As the latter records spin flips it measures the abrupt change in the number of Dirac strings threading the coil. These step-function events as shown in Figs. (1b) and (\ref{Stanford-abs}) are measurable in simulations and are related to magnetic correlations in spin ice \cite{Jaubert2011,Ryzhkin2013,Dusad2019} but have no physical meaning in the theory of monopoles.

\begin{figure}[tp]
\includegraphics[width = .9\linewidth]{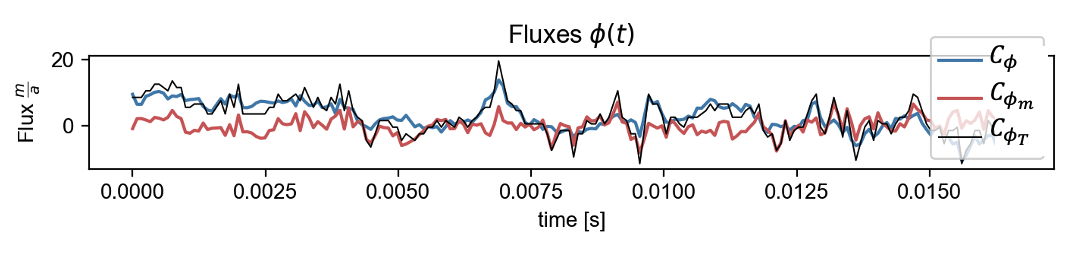}
\includegraphics[width = .9\linewidth]{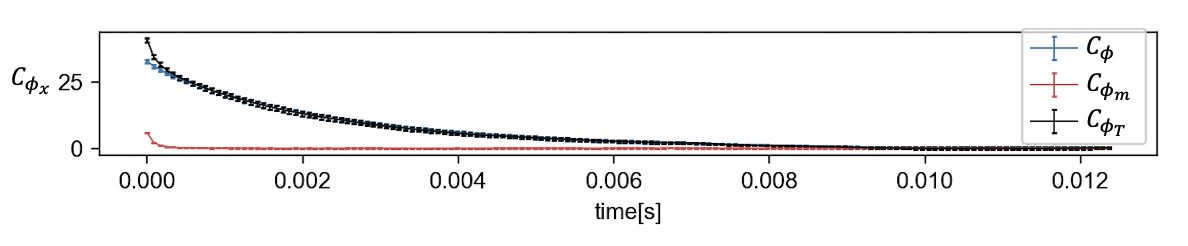}
 \caption{ {(\bf Upper}) Simulated flux through the coil against time.  {\bf(Lower)} Flux correlation functions against time - see text. Blue, magnetic field flux $\phi$, black, flux from local fields $\phi_T$, red flux from monopole alone, $\phi_m$.} \label{fluxnoise-t}
\end{figure}

\begin{figure}[tp]
\includegraphics[width = .72\linewidth]{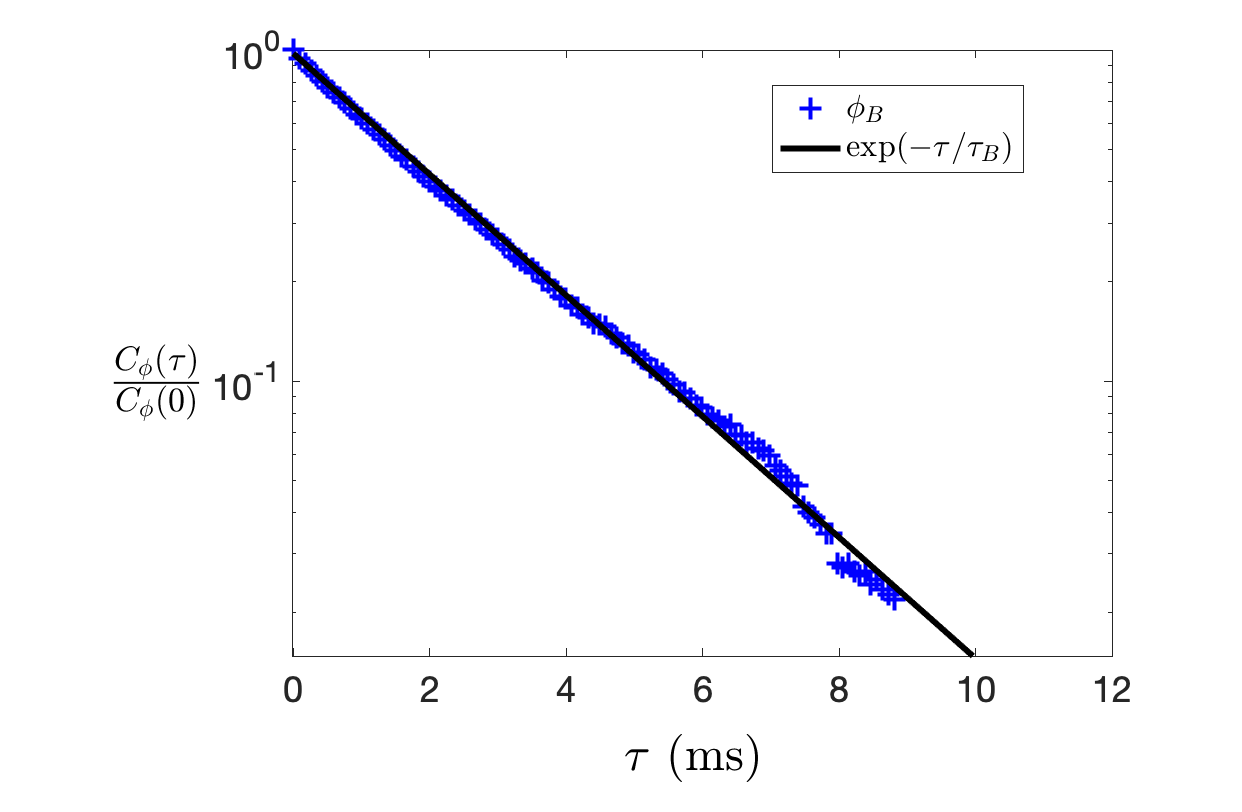}
\includegraphics[width = .65\linewidth]{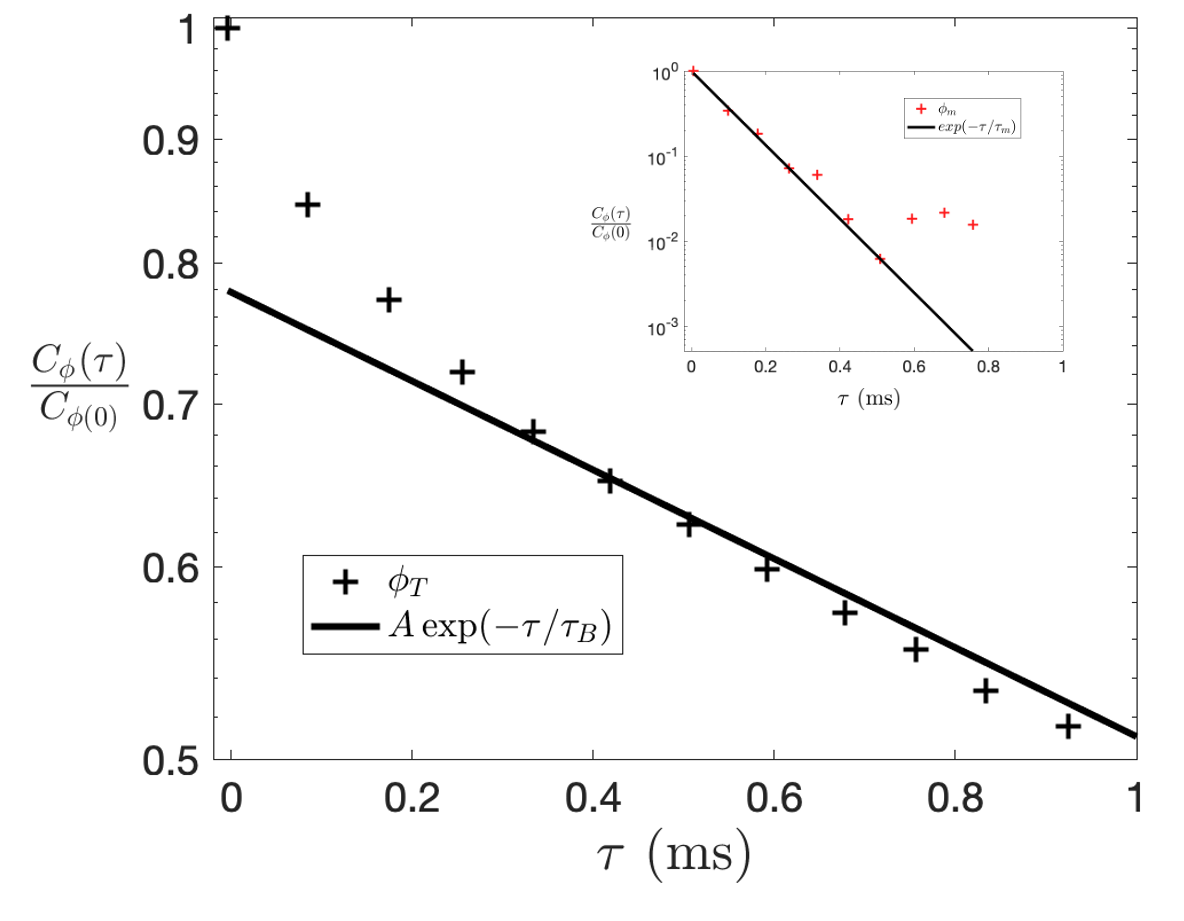}
 \caption{ {(\bf Upper}) Normalised correlation function $C_{\phi}(t)$ for the magnetic field flux $\phi$ and fit to an exponential decay, $\tau_B\approx 2.38$ ms. {(\bf Lower)} Normalised correlation function $C_{\phi}(t)$ for the  for local field flux $\phi_T$. Solid line shows an exponential decay with same time constant $\tau_B$. Inset: normalised correlation function for the flux from the monopoles, $\phi_m$ and exponential decay with $\tau_m\approx 0.1$ ms.} \label{fluxnoise-t2}
\end{figure}

\begin{figure}[tp]
\includegraphics[width = .48\linewidth]{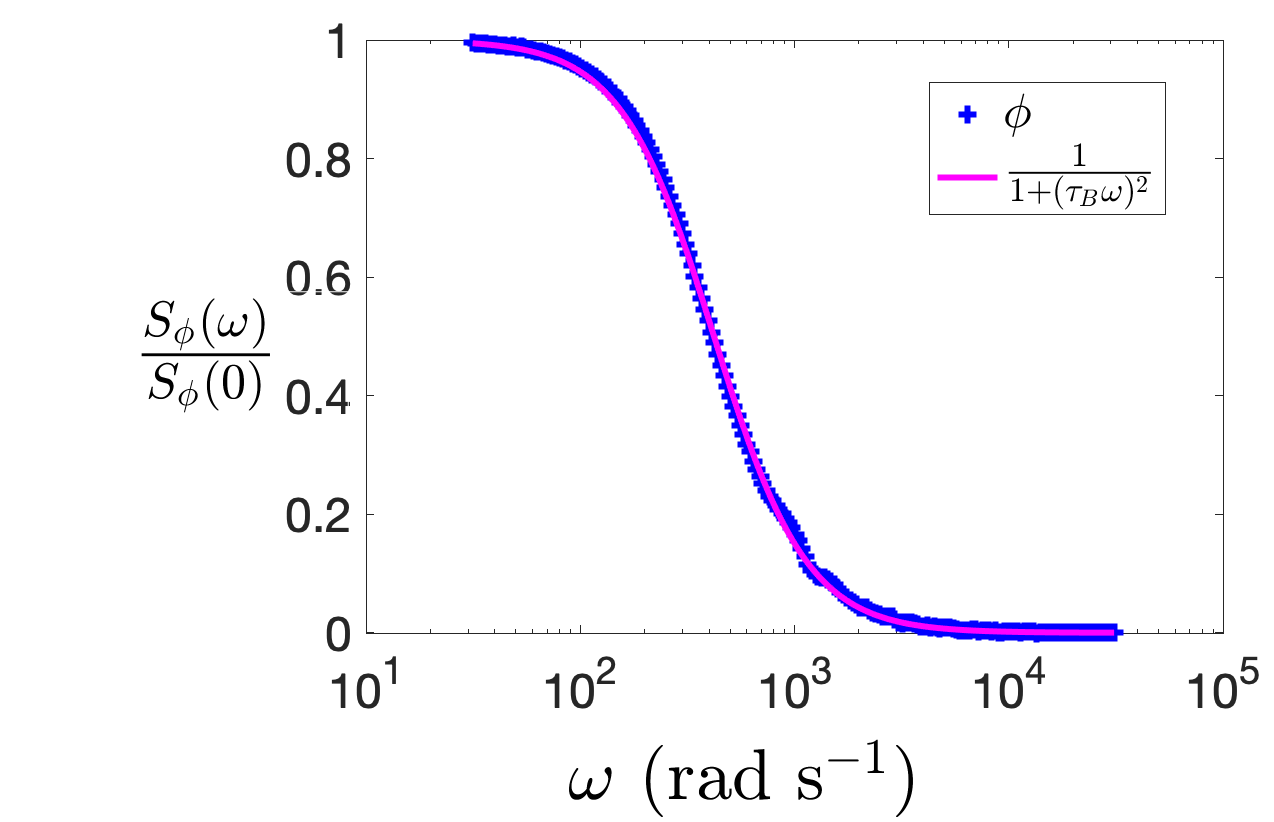}
\includegraphics[width = .48\linewidth]{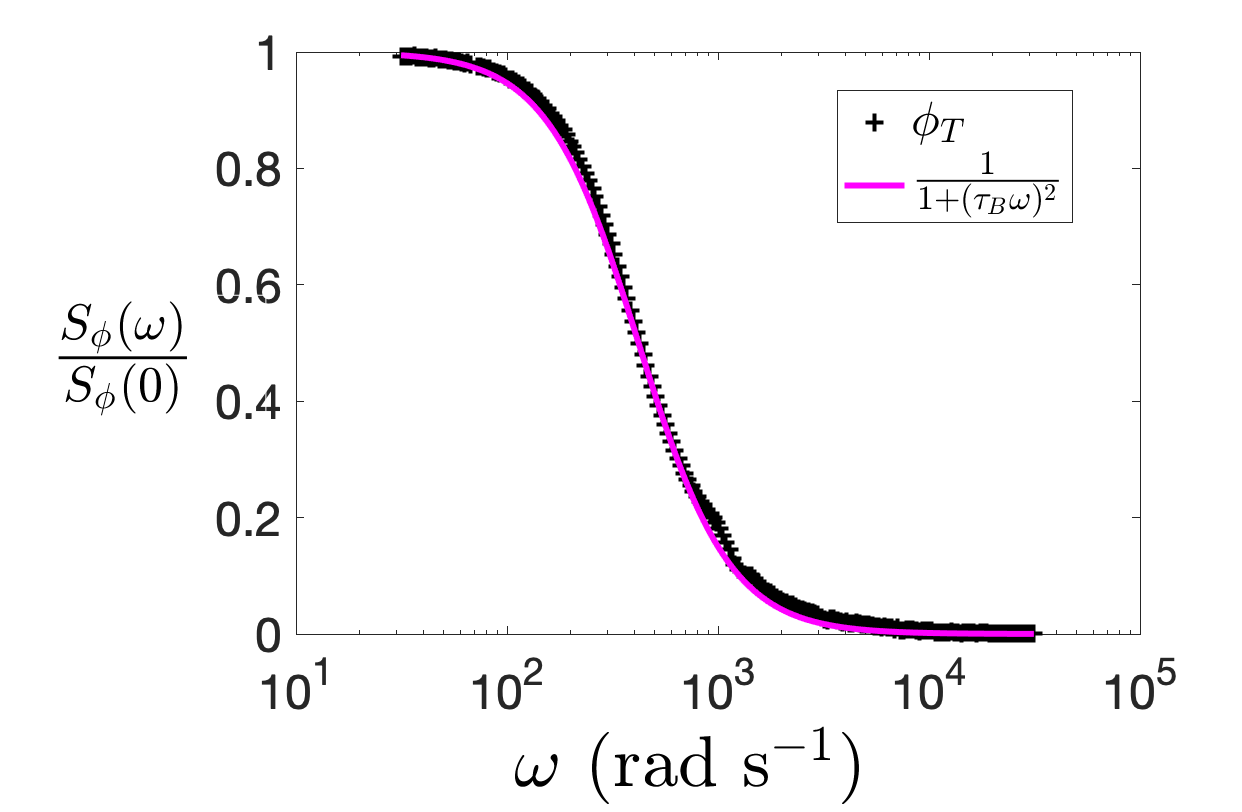}
 \caption{ {(\bf Left}) Normalised spectral density $S(\omega)$ {\it vs.} $\omega$ for magnetic field flux $\phi$. Solid line shows fit to eqn.(\ref{S-fit}) with $b=2$. {(\bf Right}) Normalised spectral density $S(\omega)$ {\it vs.} $\omega$ for local field flux $\phi_T$. Solid line shows fit to eqn.(\ref{S-fit}) with $b=2$. } \label{fluxnoise-omega}
\end{figure}

In the upper panel of Fig. (\ref{fluxnoise-t}) we show a Monte Carlo time sequence converted to milliseconds for the three fluxes: $\phi(t)$ and $\phi_T(t)$ appear closely correlated, with events stretching over a few milliseconds while $\phi_m(t)$ is characterised by shorter time scales. This is confirmed in the lower panel where we show the temporal correlation functions for the three fluxes
\begin{equation} 
C_{\phi}(t)=\langle{\int^{t_f}_0\phi(t^{\prime})\phi(t^{\prime}+t)dt^{\prime}}\rangle,
\end{equation}
where $\langle \dots\rangle$ is a thermal average and where $t_f =12$ {\it ms}, above which the signal is lost in the noise. Correlations for the magnetic field flux, $\phi(t)$ indeed fall smoothly to zero on the time scale of a few {\it ms} while $\phi_m(t)$ plunges on a much shorter time scale. It appears therefore that the Dirac strings tethered to the monopoles influence not only the amplitude of the detection signal but also the flux correlations over long times.
Correlations in the total flux $\phi_T(t)$ combine both these features starting above those for ${\phi}(t)$ at short times, falling rapidly in step with it beyond the correlation time for $\phi_m(t)$. 

Fig. (\ref{fluxnoise-t2}) show the logarithm of the normalised correlation functions against time. The upper panel shows that correlations for the magnetic field flux follow a well defined Arrhenius law within the accuracy of our simulation, with characteristic time scale $\tau_B=2.38 \pm 0.03$ ms. The lower panel shows the short time behaviour for $\phi_T$ correlations. These relax onto and become indistinguishable from the $\phi$ correlations after a few tenths of a microsecond (three or four Monte Carlo steps per spin) and beyond this time scale can be fitted with the same Arrhenius law. The inset shows the rapid fall in the $\phi_m$ correlations which fit reasonably well to an exponential decay with time constant $\tau_m \approx 0.1$ ms although a more detailed study is required to make a quantitative analysis.  

Similar relaxation with short and long time scales has previously been reported for spin correlations in the dipolar spin ice and dumbbell models \cite{Jaubert2011}. Although the quantities studied were bulk quantities they are closely related to the $\phi_T$ correlations presented above. Magnetic relaxation times in spin ice have also been estimated theoretically using an effective field approximation, equivalent to Debye-Huckel theory to treat the monopole contribution \cite{Ryzhkin2013}. Two time scales emerge from this analysis which can be interpreted as due to the Dirac strings and the monopoles. The ratio of time scales is estimated to be $9/T$ with $T$ in Kelvin, in qualitative agreement with the factor of around $20$ that we observe. It is surprising to find that the field flux correlations have the most convincing fit to an exponential. Our data suggests that the two time scales for the $\phi_T$ correlations are generated by the discrete changes in the local fields which terminate the Dirac strings. These appear smoothed out by the field variations and the emergence of continuous symmetry. 

Experimental access to magnetic noise in spin ice is principally via the spectral density \cite{Morineau2025,Samarakoon2022,Dusad2019} which we can analyse through the Fourier transform of the correlation function
\begin{equation}
S_{\phi}(\omega)=\int^{\infty}_{0}C_{\phi}(\tau)\cos(\omega\tau)d\tau.
\end{equation}
We found that this procedure left a white noise component \cite{Dusad2019} of about $2\%$ of the total signal at and above the Nyquist frequency, $\omega_{Nyq}=\frac{2\pi}{t_{nyq}}$ which was subtracted before normalisation to give the data presented in Fig. (\ref{fluxnoise-t}). The left panel shows the spectral density for the correlations of magnetic field flux $\phi$ and the right for the local field $\phi_T$. Experimental and some simulation data has been fitted to a Cole - Cole form
\begin{equation}
S_{\phi}(\omega)=\frac{S(0)}{1+(\omega\tau_B)^b}, \label{S-fit}
\end{equation}
with exponent $b$ a fitting parameter. An Arrhenius decay process with single relaxation time $\tau_B$ leads to a Lorenzian fitting function with $b=2$. It is perhaps no surprise to find that our data is well fitted with $b=2$ and the same decay time, $\tau_B\approx 2.38$ ms although we remark that the fit is quite sensitive to the value of $\tau_B$ and the goodness of fit is a reassuring consistency test. This is true for both sets of data despite the short time non-Arrhenius decay for the $\phi_T$ correlations. It is possible that a more intense numerical simulation \cite{Samarakoon2022} would find small corrections to this fit. The closeness to Arrhenius decay and fit to $b=2$ is in contrast to experiments which have found $b$ varying between $1.5$ and $2$ \cite{billington2024,Morineau2025,Samarakoon2022,Dusad2019} but in line with previous numerical work which finds $b$ around $b=1.8$ \cite{Dusad2019} and very close to $b=2$ \cite{Kirschner2018,Samarakoon2022}.


\section{Discussion}

Comparing our results with those from experiment and from simulations of realistic models of spin ice \cite{billington2024,Morineau2025,Samarakoon2022,Dusad2019,Kirschner2018} the bottom line is: they are pretty similar. This clearly shows that it is consistent to discuss the experiments in terms of monopoles and Dirac strings. The present paper illustrates the remarkable analogy between the predicted monopoles and strings of Dirac \cite{Dirac1931,Dirac1948} and the quasi-particles in model spin ice. We show that this analogy can be taken right to comparison with results from magnetic noise experiments.

This said there are a number of caveats and points of reflection to take into account. We have constructed a Faraday experiment with a coil of microscopic width, less than an Angstrom in fact. Experiment on the other hand is limited to macroscopic coils so that the measurements are bulk measurements. In such set ups creation and annihilation events \cite{Klyuev2019} occur inside the measurement device whereas in our case they are exterior to the device by construction. Experiment and certain simulations \cite{Kirschner2018} have open boundaries with surface charge. The effects of these differences could be investigated in future simulations. Direct measurement of magnetic fluctuations relates more to fluctuations in $\phi_T$ than $\phi$. It would be interesting to see if the differences between them are maintained as one changes from flux measurement to a bulk measurement. 

Finally, as exposed in recent simulations \cite{Samarakoon2022}, the complex dynamics resulting in $b<2$ is a property of spin ice, depending on the microscopic details of spin flip dynamics where one expects a large range of microscopic time scales \cite{Tomasello2019,Bovo2012} and dynamical fractal structures \cite{Hallen2022}, rather than on the constraints imposed by the presence of Dirac strings. However, we have shown that Dirac strings, or rather their termination are the key elements in the development of long time correlations in spin ice. They are the main contribution to the noise signal and their presence here, when $b=2$ to an excellent approximation shows explicitly that $b<2$ is not a prerequisite for their presence. 



\section{Conclusion}

In conclusion we have shown that the magnetic field distributions for monopole quasi-particles in model spin ice closely follow those for Dirac monopole in the vacuum. The analogy  includes classical analogues of Dirac strings that impose the divergence free condition for the magnetic field they produce. The Dirac strings are defined in space-time rather than in individual configurations and they appear in transition graphs between final and initial configurations resembling particle trajectories through the vacuum. A long string cannot be observed in an induction experiment but its termination on the monopole is an integral part of the monopole signal. We have shown that monopole noise under these conditions is very similar to magnetic noise in spin ice materials. 

In the vacuum the search for magnetic monopoles continues. Our analysis does not add directly to this quest but it does show that powerful analogies exist elsewhere. As a consequences, as recently suggested \cite{Klyuev2024}, if one could search for them in a high density environment such as in the first instances of the universe, their signal could resemble that of magnetic noise in spin ice materials. 

However, despite the closeness of the analogies presented here, the magnetic monopoles and classical Dirac strings of spin ice are not the real thing. Here magnetic fields, observable in experiment ride on the back of the emergence of classical nearest neighbour models in the absence of real dynamics. This could be corrected for in quantum spin ice \cite{Gingras2014,Savary2017} where complete quantum electrodynamics emerges with a spin liquid vacuum from which magnetic and electric charges can be excited. Coulomb interactions and real magnetic flux will be induced through quantum fluctuations of the vacuum giving access to complete monopole current and Dirac string dynamics, to monopole detection in a Stanford experiment and to real monopole noise in a dense fluid environment.
 

\begin{acknowledgements}

We thank Steven Bramwell, Rodolfo Borzi, Ludovic Jaubert and Lucile Savary for useful discussions. This work was supported by ANR grant No. ANR-19-CE30-0040 and in part by grant NSF PHY-2309135 to the Kavli Institute for Theoretical Physics (KITP). AHZ acknowledges financial support from the ENS de Lyon.

\end{acknowledgements}


 

\appendix

\section{Maxwell's equations with Dirac Monopoles}\label{Maxwell}

In the presence of monopoles Maxwell's equations must be modified to give symmetry between electric and magnetic charge:
\begin{eqnarray}
\vec \nabla. \vec E&=& \frac{\rho_e}{\epsilon_0}, \; \vec \nabla. \vec B= \rho_m  \label{Max1}\\
\curl \vec B - \frac{1}{c^2}\frac{\partial \vec E}{\partial t} &=&\vec j_e, \; -\curl \vec E - \frac{\partial \vec B}{\partial t} = \vec j_m, \nonumber 
\end{eqnarray}
where quantities have their usual meaning and $\rho_m$ and $\vec j_m$ are the monopole density and current density respectively. 

However, it is possible to transform the equations back to their unmodified form if the magnetic field is redefined to include Dirac strings lying along the monopole trajectories. The field at position $\vec r$ and time $t$ associated with the Dirac strings can be written
\begin{equation}
\vec B_{Dirac}(\vec r,t)=\sum_m \mu_0 Q_m\left[\int^{\vec r_m(t)}_{-\infty}\delta(\vec s_m-\vec r)d \vec s_m\right],
\end{equation}
where $\vec r_m(t)$ is the position of a monopole of charge $Q_m$ at time $t$, $[s_m(t^{\prime})]$ is the trajectory of points visited by the monopole during the time interval $-\infty < t^{\prime}< t$ and $d \vec s_m$ lies along the trajectory in the direction of propagation. Taking the time and spatial derivatives of $B_{Dirac}$
\begin{eqnarray}
\frac{\partial \vec B_{Dirac}}{\partial t} &=& \sum_m Q_m\vec v_m \delta(\vec r_m(t)-\vec r)= \vec j_m, \nonumber \\
\vec \nabla .\vec B_{Dirac} &=& -\sum_m Q_m \delta(\vec r_m(t)-r)= -\rho_m.
\end{eqnarray}
Re-defining $\vec B^{\prime}=\vec B_{Dirac}+\vec B$ it follows that:
\begin{eqnarray}
\vec \nabla. \vec E&=& \frac{\rho_e}{\epsilon_0}, \; \vec \nabla. \vec B^{\prime}= 0  \label{Max2}\\
\curl \vec B^{\prime} - \frac{1}{c^2}\frac{\partial \vec E}{\partial t} &=&\vec j_e, \; -\curl \vec E - \frac{\partial \vec B^{\prime}}{\partial t} = 0. \nonumber 
\end{eqnarray}

For a single monopole in the vacuum, identifying $\vec B$ with $\vec B_m$ we recover eqn. (1) for the Stanford experiment. Faraday's law, found from the fourth Maxwell equation, is modified by a contribution from the monopole current  in eqns. (\ref{Max1}). It remains unchanged in the Dirac string picture (eqns.(\ref{Max2})) but now the Dirac string makes an additional contribution to the time dependence of the field. 
The contributions to the signal in Stanford experiment are identical and the two forms are equivalent.
The advantage of the Dirac string construction for analogies with spin ice is clear although it gives a physical sense to the Dirac string over and above that emerging from Dirac's theory \cite{Dirac1931,Dirac1948}. It also puts constraints on the gauge invariance. To define the Dirac string in space the vector potential must be set to zero along its length whereas in the first formulation it is only fixed at the instantaneous monopole positions giving more gauge freedom. 

\section{Fragmentation of a spin ice configuration}\label{Frag-App}

We have used the iterative real space algorithm proposed by Slobinsky {\it et. al.} [\onlinecite{Slobinsky2019}] to decompose the set of elements $M_{IJ}=M^m_{IJ}+M^d_{IJ}=\pm 1$ in units of $\frac{Q}{2}$ into their fragmented elements $M_{IJ}^d$ and $M_{IJ}^m$. The four fluxes associated with tetrahedron $I$ are $M_I=[M_{IJ},M_{IK},M_{IL},M_{IM}]$ with charge $Q_I = -\sum_J{M_{IJ}}$.
The algorithm  decomposes the four-vectors iteratively, converging to a solution where the $d$ component is divergence free: $Q^d_I=\sum_J{M^d_{IJ}}=0 \; \forall \; I$ so that eqn.(\ref{divf}) is satisfied. 

We take the example of an isolated north pole on tetrahedron $A$ 
with initial configuration $[-1,-1,-1,1]$. Its four neighbouring sites satisfy the ice rules; one of which, $B$ has configuration $[1,1,-1,-1]$, where $M_{AB}=-M_{BA}=-1$ is the common needle. For the zeroth order iteration we set the monopolar fields to zero giving starting configurations:
\begin{align}
\begin{split}
    M^0_A = [-1,-1,-1,1]_d + [0,0,0,0]_m \\
    M^0_B = [1,1,-1,-1]_d + [0,0,0,0]_m
\end{split}
\end{align}
The object is to exchange flux (charge) from the $d$ to the $m$ sector. The initial charges in the $d$ sector are $Q^d_A=2$ and $Q^d_B=0$. To exchange flux we note that $M^m_{IJ}$ depend on the difference in charge on the connected sites and define a flow of charge from $d$ to $m$
%
\begin{equation} \label{eq:flow}
    F_{AB} = \frac{Q^d_A-Q^d_B}{8} = \frac{1}{4},
\end{equation}
where the factor $\frac{\Delta Q}{8}$ is $\frac{\Delta Q}{2}$ divided equally between the four needles of $A$ \cite{Slobinsky2019}.  The re-defined fluxes for the next iteration, which conserve the total charge on each tetrahedron are:
%
\begin{align}
\begin{split}
    \label{eq:carryfluxes}
    A-B: \enspace M_d^1 = M_d^0 + F_{AB}, \; \; M_m^1 = M_m^0 - F_{AB} \\
    B-A: \enspace M_d^1 = M_d^0 - F_{AB}, \; \; M_m^1 = M_m^0 + F_{AB}.
\end{split}
\end{align}
%
This is repeated for the three other bonds of tetrahedron $A$, using in each case the fluxes at level $0$. In this example all the $F_{IJ}=1/4$, however if the $\Delta Q$ for the four neighbouring tetrahedra are different, the flux exchanges for the four bonds will be different. Fluxes after the first iteration are then:
%
\begin{align}
    M^1_A = [-\frac{3}{4},-\frac{3}{4},-\frac{3}{4},\frac{5}{4}]_d + [-\frac{1}{4},-\frac{1}{4},-\frac{1}{4},-\frac{1}{4}]_m
\end{align}

In the second iteration, we again calculate the fluxes $F_{IJ}$ given by the $Q^d$ charges in equation \ref{eq:flow} and apply them to the fields for the corresponding needles using \ref{eq:carryfluxes}. In this iteration there is exchange from the $d$ to the $m$ sector at sites $C$ that are second neighbours to $A$. The iteration is continued until a given threshold $|Q_I^d| < \lambda$ is reached for each site.

In our example we converge to the solution:
\begin{align}
    M^n_A = [-\frac{1}{2},-\frac{1}{2},-\frac{1}{2},\frac{3}{2}]_d + [-\frac{1}{2},-\frac{1}{2},-\frac{1}{2},-\frac{1}{2}]_m,
\end{align}
as in eqn.(\ref{frag-simple}).
Solving for the whole system simultaneously, the monopolar field lines are propagated outwards (Fig. \ref{flux-phim}) and are non-zero even far from the monopoles, creating long-range correlations and ensuring that $M^m$ is also divergence free everywhere except at the sites containing monopoles.

We typically imposed $\lambda=0.01$ which was reached everywhere for $n \approx 60$ iterations for a system size of $L=8$.

\bibliography{biblio-mono}

\end{document}